\documentclass[12pt]{article}

\usepackage{amsmath, amssymb,upgreek}
\usepackage[dvips]{graphicx}

\begin{document}
\begin{center}
{\bfseries \uppercase{On asimuthal anisotropy in fragmentation\\ of classical relativistic string}}

\vskip 5mm

R.S. Kolevatov\footnote{E-mail: rodion.kolevatov@fys.uio.no}

\vskip 5mm

{\small {\it
Department of Physics, University of Oslo, PB1048 Blindern, N-0316 Oslo, Norway \\
{\rm on leave of absence from}\\
Department of High Energy Physics, Saint-Petersburg State University,\\  Ulyanovskaya 1, 198504 Saint-Petersburg, Russia
}}
\end{center}

\vskip 5mm

\begin{center}
\begin{minipage}{150mm}
\centerline{\bf Abstract}
A fragmenting relativistic string is widely used for modelling particle production via quark-gluon strings formed in hadron inelastic interactions of high energies. In this note we focus on motion and fragmentation of relativistic string with non-zero transverse separation of its ends and study this scenario as a possible mechanism bringing anisotropy into the asimuthal angle disribution of produced particles in inelastic interactions of hadrons. 
\end{minipage}
\end{center}

\vskip 10mm

\section{Introduction.}
A two-stage scenario of particle production through the formation and subsequent break-up of quark-gluon strings via Schwinger mechanism is a common picture for inelastic hadronic interactions \cite{Kaidalov-QGSM,Capella-QGSM}. A widely used approximation for quark-gluon string is the classical relativistic string with break-ups described by the area law \cite{Artru}. General picture used in a number of models of particle production via quark-gluon string decay (e.g. \cite{Sjostrand-PYTHIA,Werner-VENUS}) assumes that a quark-gluon string starts stretching from a single point in space though may still have a complicated structure with kinks in the middle. However, this picture may yet capture not all of the features which may be important for some of the observables. Namely, constituents of interacting hadrons (valence quarks and diquarks, sea $q \bar q$ pairs) are at different positions in the transverse plane during the interaction with a typical separation determined by the value of inelastic cross section, that is, around 1~fm, which is considerably less than typical string transverse size. 

Another hint comes from the recent studies \cite{BKK}, where it was shown that asimuthal anisotropy in $p_t$ distribution is observed in Gribov-Regge theory for particles produced via single Pomeron exchange if the constituents of the incoming hadrons or nuclei, which the Pomeron is attached to, have different positions in the transverse plane with the preferred in-plane (i.e. parallel to the overall impact parameter) orientation. In hadron-hadron and nucleus-nucleus collisions this in-plane orientation appears to be prevailing due to the gradients of the optical thicknesses of the colliding systems. At the same time the cut Pomeron is usually identified with a couple of quark-gluon strings \cite{Kaidalov-QGSM}. So, in present study we address a question whether fragmentation of a classical relativistic string with a certain transverse separation of its ends could produce asimuthal anisotropy in transverse momentum distribution of the fragmentation products. We consider motion and fragmentation of a classical relativistic string with initial transverse separation of its ends of the order of $b=1$~fm.

The structure of the paper is as follows.
In two sections following introduction we review generalities of string motion which are well-known and make some general notes which are relevant for our subsequent numerical calculations. The forth section is devoted to motion of a relativistic string starting from some two special initial conditions. In the fifth section we describe a procedure which we use for string fragmentation in our Monte-Carlo code. Numerical results for two types of initial conditions described in section 4 are presented in subsequent section. Then follows a conclusion and an outlook.

\section{Relativistic string, equation of motion.}
We start with reviewing the basics of string motion and fragmentation. A detailed derivation of the equations of motion starting from the string Lagrangian can be found elsewhere (see e.g. \cite{Artru} or \cite{Werner-VENUS} and references therein). Here we just list the results which we shall use in subsequent consideration.

In general, coordinates of the points of the string $x^\mu$ are functions of two parameters, the timelike, $\tau \in (-\infty, \infty)$ and the spacelike $\sigma$ (which takes values in a limited interval). To constrain freedom in reparameterization one imposes a specific gauge on the $x^\mu$ as functions of $\sigma$ and $\tau$. For clarity and transparent physical interpretation of $\tau$ and $\sigma$ the orthonormal gauge is convenient, together with so-called lab frame parameterization. The lab frame parameterization corresponds to identifying time-like parameter $\tau$ with time: $$x^0=\tau \equiv t.$$ Using this, the orthonormal gauge reads:
\begin{equation}
 \left\{ \begin{array}{rcl}
 \dot {\bf x} \cdot {\bf x}' & = & 0,\\
 \dot {\bf x}^2 +{\bf x}'^2& =&1. 
\end{array}
\right.  \label{eq-gauge}
\end{equation}
where $\dot {\bf x} \equiv \cfrac{\partial \bf x}{ \partial t}$ and $ {\bf x}' \equiv \cfrac{\partial \bf x}{ \partial \sigma}$. In this gauge string's equations of motion are just wave equations
\begin{equation}
\ddot {\bf x} - {\bf x}''  =  0   \label{eq-motion}
\end{equation}
with boundary conditions
\begin{equation}
 \left.{\bf x}'\right|_{\sigma=0,\sigma_{\rm max}}=0. \label{eq-boundary}
\end{equation}

Energy and momentum of a small piece of a string are
\begin{equation}
 d p^\mu = \varkappa \dot x^\mu(t,\sigma) d\sigma \quad 
 \text{or} \quad d E = \varkappa d\sigma, \quad d{\bf p} = \varkappa \dot {\bf x} d\sigma. \label{eq-energy}
\end{equation}
where $\varkappa$ stands for string tension.
So, parameter $\sigma$ itself has a transparent physical interpretation, as the conserved energy and momentum of the string are
\begin{eqnarray}
 E & = & \varkappa \int_0^{\sigma_{\rm max}} d \sigma = \varkappa \sigma_{\rm max};  \label{string:energy}\\
 {\bf p} & = & \varkappa \int_0^{\sigma_{\rm max}} \dot {\bf x} d\sigma, \label{string:momentum}
\end{eqnarray}
Hence, in the lab frame parameterization $\sigma \in (0, E/\varkappa)$.

Solution of the equations of motion (\ref{eq-motion}) which satisfies boundary condition (\ref{eq-boundary}) can be expressed as
\begin{equation} 
{\bf x} = \cfrac{1}{2} [ {\bf y}(t+\sigma) + {\bf y} (t-\sigma)], \label{string:solution}
\end{equation}
where ${\bf y}(t)$, a function of single variable, is obviously the trajectory of one of the string's end points. This trajectory is usually called directrix. In terms of the directrix, gauge conditions (\ref{eq-gauge}) are equivalent to $${ \bf y'}^2 = 1.$$ Momentum conservation (\ref{string:momentum}) implies that directrix is ``periodic'':
\begin{equation}
 {\bf y}(t+ 2 \sigma_{\rm max}) - {\bf y}(t) = 2{\bf p} / \varkappa.
\end{equation}
This means, we need to know the trajectory of one of the string's ends on a finite time interval to be able to fully describe string's motion. At the same time a full picture of the relativistic string motion is unambigously defined, if we know initital conditions, that is, momenta and arrangement of string at some fixed moment. 

We devote the next section to linking these two points.

\section{Initial conditions and directrix recovery.}
As it has already been mentioned in the introduction, to the best of our knowledge, most applicatons use initial conditions for the relativistic string as a source of particles which assume that string is stretched from a single point in space.
In this case, subsequent string motion is fully described once we know momenta of string's ends and kinks at the time of its creation.
However, it seems natural to consider situations with non-zero length of a string at the initial stage of the evolution (non-zero transverse separation for the string's ends)  with some momentum distribution along the string. 
One special case of motion of a relativistic string with non-zero ends' separation was considered in \cite{Peschanski}. Approximation used there assumes string's ends moving at constant velocity $v<c$ which physically implies their masses being infinite. Contrary, here  we shall consider a string with massless ends as a usual approximation used in models of hadron production in inelastic interactions. 

We proceed with the technical issue of recovering directrix from coordinates and momenta of strings pieces at a given time $t_0$ in a way which can be used in numerical simulations. 
Let us assume that for this time moment $t_0$ we are provided information on string arrangement in space and momentum distribution along the string length $l$, ${\bf x}(l)$ and $\tfrac{d{\bf p}}{dl}(l)$.
These initial distributions cannot be arbitrary, the restrictions can be read out from the gauge conditions (\ref{eq-gauge}). One can fulfill gauge conditions having rather ${\bf x}'\ne 0$ or ${\bf x}' = 0$. The first opportunity corresponds to a smooth distribution of $\frac{d\bf p}{dl}$ with $\tfrac{d \bf x}{dl} \perp \tfrac{d\bf p}{dl}$. The second opportunity, ${\bf x}'=0$ for a certain range of parameter $\sigma$, implies $|\dot {\bf x} |=1$ with arbitrary direction for $\dot {\bf x}$ for that $\sigma$ range. This corresponds to a 'kink' situation where part of a string of exactly zero length in space carries a certain finite momentum leading to a delta-function-like contribution to the $d{\bf p}/dl$. 

The second of the gauge equations gives a relation expressing increment of the parameter $\sigma$ along the infinitezimal piece of a string in terms of its length $dl=|d\bf x|$ and momentum $d\bf p$: 
\begin{equation}
d\sigma = dE / \varkappa = \sqrt{d{\bf p}^2+\varkappa^2 d l^2}/\varkappa, \label{string:dsigma}
\end{equation}
which upon integration provides dependence $l(\sigma)$. This gives ${\bf x}'$ and $\dot {\bf x}$ as functions of $\sigma$ for a given time $t_0$:
\begin{equation}
 \left\{ \begin{array}{rcl}
          {\bf x}' &=& \cfrac{d\bf x}{d l}\cfrac{\varkappa}{\sqrt{\left({d\bf p}/{dl}\right)^2+\varkappa^2}},\\
          \dot{\bf x} &=& \cfrac{d\bf p}{d l}\cfrac{1}{\sqrt{\left({d\bf p}/{dl}\right)^2+\varkappa^2}};  
         \end{array}
\right.\quad
\text{ and } \quad 
\left\{ \begin{array}{rcl}
          {\bf x}' &=& 0,\\
	  \\
          \dot{\bf x} & = & \cfrac{1}{\varkappa}\cfrac{\Delta {\bf p}_{\text{kink}}}{\Delta \sigma_{\text{kink}}},\; 
	  \Delta \sigma_{\text{kink}} =\cfrac{|\Delta {\bf p}_{\text{kink}}|}{\varkappa} ;  
         \end{array}
\right. \label{xdotprime}
\end{equation}
for smooth part and kinks respectively.

At the same time, as follows from (\ref{string:solution}),
\begin{eqnarray}
&&{\bf x}'(t_0, \sigma)  =  \frac{1}{2} \left[ {\bf y}' (t_0+\sigma) - {\bf y}' (t_0-\sigma)\right]; \label{string:xprime}\\
&&\dot {\bf x}(t_0, \sigma)  = \frac{1}{2} \left[ {\bf y}' (t_0+\sigma) + {\bf y}' (t_0-\sigma)\right] . \label{string:xdot}
\end{eqnarray}
So 
\begin{equation}
\begin{array}{l} 
{\bf y}'(t_0+\sigma) = \dot {\bf x}(t_0, \sigma) + {\bf x}'(t_0, \sigma),\\
{\bf y}'(t_0-\sigma) = \dot {\bf x}(t_0, \sigma) - {\bf x}'(t_0, \sigma).\\
\end{array}
\label{string:increments} 
\end{equation}
Upon integrating (\ref{string:increments})  and taking into account that ${\bf y}(t_0) = {\bf x}(t_0, \sigma = 0)$ is a position of one of the string's ends, one recovers the directrix on the interval $ t \in (t_0-\sigma_{\rm max},t_0+\sigma_{\rm max})$, and, hence, at any point due to its periodicity. These equations can be also written in a form convenient for numerical applications, in terms of momenta and lengths of infinitesimal string pieces:
\begin{equation}
\begin{array}{l} 
d{\bf y}(t_0+\sigma) \equiv {\bf y} (t_0+\sigma+d\sigma) - {\bf y}(t_0+\sigma) = (d {\bf p}/\varkappa + d\bf x),\\
d{\bf y}(t_0-\sigma) \equiv {\bf y} (t_0-\sigma-d\sigma) - {\bf y}(t_0-\sigma) = -(d {\bf p}/\varkappa - d\bf x). 
\end{array}
\label{string:increments1} 
\end{equation}

\section{Examples of motion.}

To illustrate the technique described in the previous section, we reconstruct directrix for two simple model initial conditions. These are in some sence extreme cases which implement two opposite possibilities, namely $a)$ rapidity gap with large transverse separation between partons at strings ends and $b)$ smooth distribution of momentum carried by string pieces with linear relation between position of a piece in the transverse plane and its rapidity $y$.

\begin{figure}[h]\centering
\includegraphics[width=0.3\hsize]{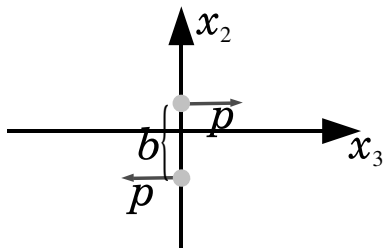}  \caption{'Kink-type' initial conditions.}  \label{string:fig:init1}
\end{figure}

We start with the case $a)$. Consider a string with massless ends separated by impact parameter $\bf b$ along the $x_2$-axis connected by a string piece being initially at rest.  Ends of the string carry momenta $p$ and $-p$ along the $x_3$ as shown in fig.~\ref{string:fig:init1}. We call this 'kink-type' initial conditions in what follows. In this example total energy of the string is $E = 2 p + \varkappa b$, and $\sigma\in (0, 2p/\varkappa+b)$. The motion of a string with this kind of initial conditions will evidently be two-dimensional with trajectory lying in $(x_2,x_3)$ plane.

To recover directrix we'll start with drawing a trajectory of the upper endpoint of the string for $(0,\sigma_{\max})$ time interval, that is using first line of (\ref{string:increments1}). At zero time the endpoint carries momentum $p$ with length attributed to endpoint being zero. So for $\sigma \in (0, p/\varkappa)$ we should use (\ref{string:increments1}) setting $d{\bf x}=0$ while $d{\bf p}/\varkappa$ differential is non-zero and parallel to $x_3$ axis. For $\sigma \in ( p /\varkappa,  p / \varkappa +b)$, we have only the second one of the two differentials in the directrix increment (\ref{string:increments1}) which goes against $x_2$. For $\sigma \in (p/\varkappa + b, 2 p/\varkappa +b)$ again the first differential plays and the increment goes against $x_3$.
Directrix pieces for $t\in(-\sigma_{\max},0)$ are constructed in exactly the same manner making use of the second line of (\ref{string:increments1}).

\begin{figure}[!h]
 \centering 
{({\it a})\includegraphics[width=0.9\hsize]{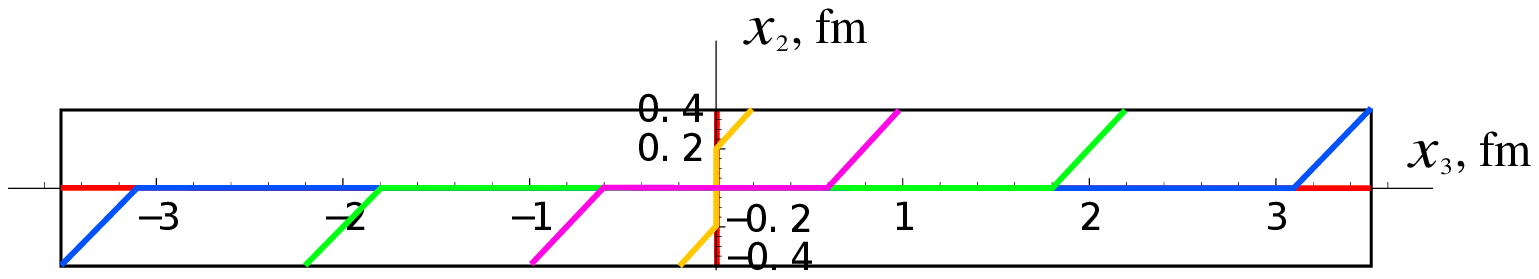} \phantom{({\it a})}}
\vskip 1cm
{({\it b})\includegraphics[width=0.9\hsize]{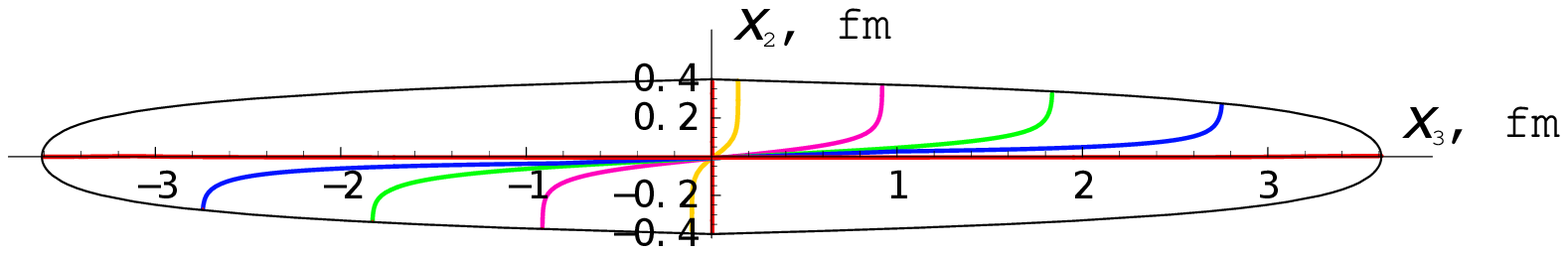}\phantom{({\it b})}}  \caption{Examples of string motion with different initial conditions: (a)~-- 'kink-type', (b)~-- 'invariant-type'. For both cases impact parameter $b=0.8$~fm, sring c.m. energy $\sqrt{s}=7.4$~GeV, string tension $\varkappa = 1$~GeV/fm.} \label{string:figure:invar}
\end{figure}

Finally, combining the differentials which is straightforward, we find that the directrix in this case represents a rectangular box with argument $\sigma$ changing linearly along the rectangle lines. This means that the ends are moving along these lines with the speed of light. Folowing the prescription (\ref{string:solution}) we find position of the string at arbitrary moment. The directrix is shown in fig.~\ref{string:figure:invar}{\it a}  together with string positions at some selected time moments. Note that a transverse part does not disappear instantly and that parts of the strings which are at 45 degrees with $x_3$ axis also carry momentum which has a component along $x_2$. So, if fragmentation of string starts and two subsequent breaks occur within the inner part cutting off the string piece, it will gain an additional contribution to the momentum along $x_2$.

For a string piece with longitudinal momentum $dp_3$ and energy $dE$ one can find its rapidity according to usual definition
$$
y=\frac 1 2 \ln \frac{dE+dp_3}{dE-dp_3},
$$ or in terms of piece velocity:
\begin{equation} 
y= \frac 1 2 \ln\frac{1+\dot x_3}{1- \dot x_3}, \label{eq-piece-rapidity} 
\end{equation}
the latter definition is also suitable for a point on a string except the kinks.

In the example described above, 'kink' case $a)$, strings ends were connected by a piece of a string with exactly zero rapidity in the lab frame. 
For the second example, case~$b)$, we take a prescription for momentum distribution along the string which does not lead to appearing of some distinguished rapidity value and which is invariant with respect to the boosts along the $x_3$ axis. The possible choice which satisfies this condition is to assume that string is still a straight line in transverse plane, with rapidity, defined according to (\ref{eq-piece-rapidity}), changing linearly with the transverse position of the point. We denote the proportionality coefficient by $\alpha$. This gives for a momentum and energy of a small string piece of length $db$:
\begin{equation}
 d p_3 = \pm \varkappa db \sinh (\alpha |{\bf b} - {\bf b_c}| - y_c). \label{string:invarparam}
\end{equation}
This leads to
\begin{equation}
 d E = \varkappa db \cosh (\alpha |{\bf b} - {\bf b_c}| - y_c)
\end{equation}
due to (\ref{string:dsigma}). Here ${\bf b}_c \equiv ({\bf b}_{\rm max} + {\bf b}_{\rm min})/2$ denotes position of the string geometrical center, $y_c$ is the relative rapidity of the string's center of mass and lab frames while $\alpha |{\bf b} - {\bf b_c}|$ corresponds to the rapidity of the piece with transverse position ${\bf b}$ in the string center of mass frame. Once the string energy in its c.m.frame $\sqrt{s}$ and transverse separation of ends are given, the coefficient $\alpha$ can be extracted from
\begin{equation}
\sqrt{s} \equiv E_{\rm c.m.} = \int\limits_{-|{\bf b}_{\max}-{\bf b}_{\min}|/2}^{|{\bf b}_{\max}-{\bf b}_{\min}|/2} d b \cosh(\alpha b )=
 \frac{2}{\alpha}\sinh\frac{\alpha|b_{\max}-b_{\min}|}{2},
\end{equation}
e.g. solving it iteratively.

If the same string is considered in another reference frame boosted along $x_3$ with respect to the initial one, rapidity attributed to the string point has again a linear dependence on its transverse position with the same proportionality coefficient $\alpha$. The only thing which has to be changed in (\ref{string:invarparam}) is the value of $y_c$. In this sence this prescription for momentum distribution along the string is boost-invariant, so in what follows we refer to is as 'invariant-type' initial conditions. The distribution of longitudinal momentum (\ref{string:invarparam}) in this example coincides with what is found in \cite{Peschanski} for the string with infinite masses on its ends at zero time. 

It is also possible to write down an intermediate variant which has both mentioned types of initial momentum distributions as limiting cases. Namely one does so setting 
\begin{equation} d p_3 = \pm A \varkappa db \sinh (\alpha |{\bf b} - {\bf b_c}|) \label{init-intermediate}\end{equation}
in the string center of mass frame. $A$ is an apriori fixed constant and $\alpha$ is defined to fit a particular mass of the string; limits $A\to \infty$ and $A=1$ give the 'kink' and 'invariant' cases respectively. However it is only for $A=1$ that the argument of ``$\sinh$'' function has a meaning of rapidity of an infinitesimal string piece and that transformations to other reference frames imply just adding rapidity difference to the argument without changing ``$\sinh$'' law.

The directrix for the 'invariant' case and string position at any time can be obtained in the same fashion as in the previous example.
Let us write down the directrix for parameterization defined by $\alpha$ and impact parameter in its center of mass frame. Put center of coordinate system to the center of the string. Projections of the directrix increments (\ref{string:increments}) on $x_2$ and $x_3$ connected with the piece of a string of length $dx_2$ (string lies along $x_2$ at zero time) are
\begin{eqnarray}
 dy_2=dx_2;\\
 dy_3=dp_3/\varkappa = dx_2 \sinh (\alpha x_2).
\end{eqnarray}
Integration from $x_{2 \max}$ downto $x_2$ gives parametric dependence of the directrix on the transverse coordinate:
\begin{eqnarray}
 y_2 = x_2;\\
 y_3 = \frac{1}{\alpha}( \cosh \alpha x_2 -\cosh \alpha x_{2 \rm max}),
\end{eqnarray}
where $x_2$ takes values in $x_2\in (-x_{2 \rm max}, x_{2 \rm max})$. The last thing to note is that, the argument of the directrix is equal to its length measured from the point at which we start the directrix construction due to the condition $|{\bf y}'|=1$ .

An example of this type of directrix together with string positions at some time moments is presented in the figure \ref{string:figure:invar}~\,{\it b}.
This motion is again two-dimensional and should lead to asimuthally anisotropic distribution of fragments on decay. The motion in $(x_2,x_3)$ plane will again give additional contribution to particle momenta along the $x_2$ axis.

As the induced transverse momentum strongly depends on the transverse velocity of the string parts, the flow-like anisotropy for both cases considered should be sensitive to both the impact parameter and center of mass energy of the initial string, namely growing with impact parameter and decreasing with c.m. energy.

\section{Fragmentation model.}

To estimate induced anisotropy in asimuthal distribution of the produced particles we use the off-shell resonance model (AMOR) \cite{Artru} with simplifications (the alternative could be effective Field-Feynman approach \cite{Field:1977fa}). The AMOR model follows the fragmentation process as it is seen from the particular reference frame. At each point of the string a break probability exists. A probability of breaking within a string piece within certain time interval is proportional to the surface element of the worldsheet swept by the piece in space-time. That is break probability within string piece corresponding to $d\sigma$ interval ($g$ stands for metric on the string worldsheet) is
\begin{equation}
 \frac{dP}{d t} = k \det(-g) d\sigma = k d\sigma |{\bf x}'| \sqrt{1-\dot {\bf x}^2 } \label{string:breakprob}
\end{equation}
with $k$ being some phenomenological constant. In this picture (for a reasonable set of parameters $k$ and string tension $\varkappa$ giving $1\div2$ particles per unit rapidity) the string first breaks into the so-called {\it primary fragments} with mass $M^2 \simeq 0.5$~GeV$^2$. As breaking of these fragments is an essentially quantum process due their low mass, they are considered as resonances off the mass shell and decay (if to neglect spin effects) isotropically in their lab frame producing observed mesons (two pions in most cases) \cite{Werner-VENUS,AGGK}.

The equation (\ref{string:breakprob}) has a transparent physical interpretation, since this means the probability of breaking within some string piece is proportional to its length times the Lorentz-factor of the piece. This means in particular, that for two model cases considered in the previous section we shall have different distributions of break points in space and time and consequently different asimuthal anisotropy patterns for produced particles.

Following prescription (\ref{string:breakprob}) we make a MC code for string breaking which explicitly observes the string dynamics. The basic scheme of the code is as follows. 

1. After directrix is reconstructed numerically from the initial conditions as described in previous sections, time evolution of the string arrangement in space is done step by step. On each step decay probability is computed by integration of (\ref{string:breakprob}) along the string length, and, according to the obtained probability, it is decided whether string will break at this time step or not.

2. If the string breaks at this step, the break point is defined, again making use of (\ref{string:breakprob}). It is assumed, that massless quark-antiquark pairs emerge exactly in the break point and have zero momentum, so the energy is conserved. This assumption is the main simplification of our fragmentation model.

3. After the breakup point is generated, the two resulting string pieces are considered as independent, and are treated in the same fashion as the parent one unless the mass of the fragment is below some cutoff $m_{\rm cut}$. 

4. If the mass of a newly produced fragment is below the cutoff mass $m_{\rm cut}$, it decays isotropically in its rest frame into two $\pi$-mesons. This decay is the main source of transverse momentum of the final particles in the model (aditional contribution comes from the classical string motion), so the transverse momentum distribution is closely linked to the mass distribution of the primary fragments.

We also do not take into account different flavour and baryonic content of the string decay products.
Though the model is simplistic, to our mind it keeps the main feature relevant for our purpose, which is evaluation of asimuthal anisotropy in momentum distribution of fragmentation products. Namely, final particles obtain a random isotropic transverse momentum of the order of $\pi$-meson mass (depends on cutoff $m_{\rm cut}$) and an additional 'in-plane' contribution to the momentum, which comes from the motion of a string.

The parameters used in the MC code are string tension $\varkappa$, breaking probability $k$ (see (\ref{string:breakprob})) and cutoff mass $m_{\rm cut}$.

\section{Numerical results.}

\begin{figure}
\hfill 
\includegraphics[width=0.5\hsize]{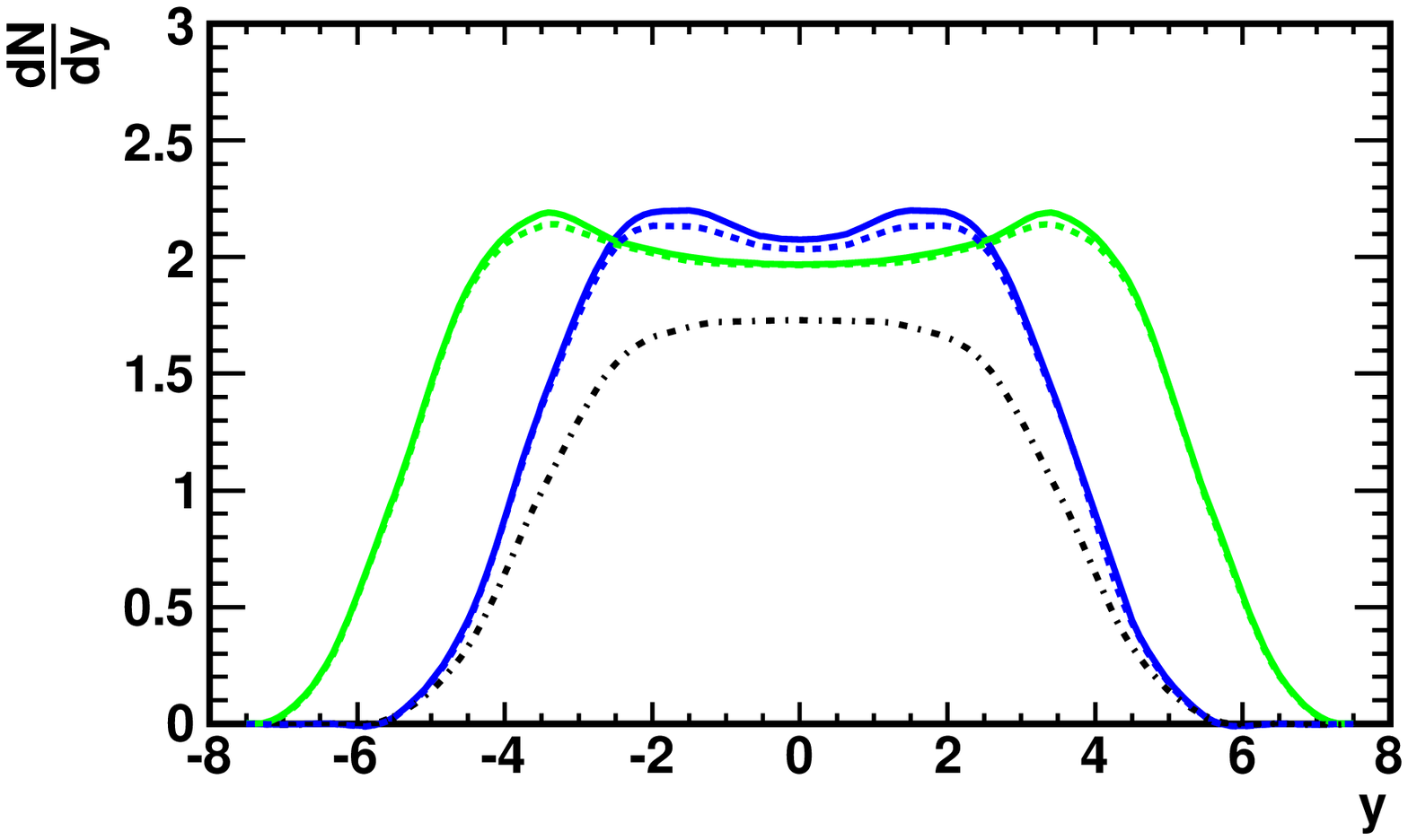}\hfill\includegraphics[width=0.5\hsize]{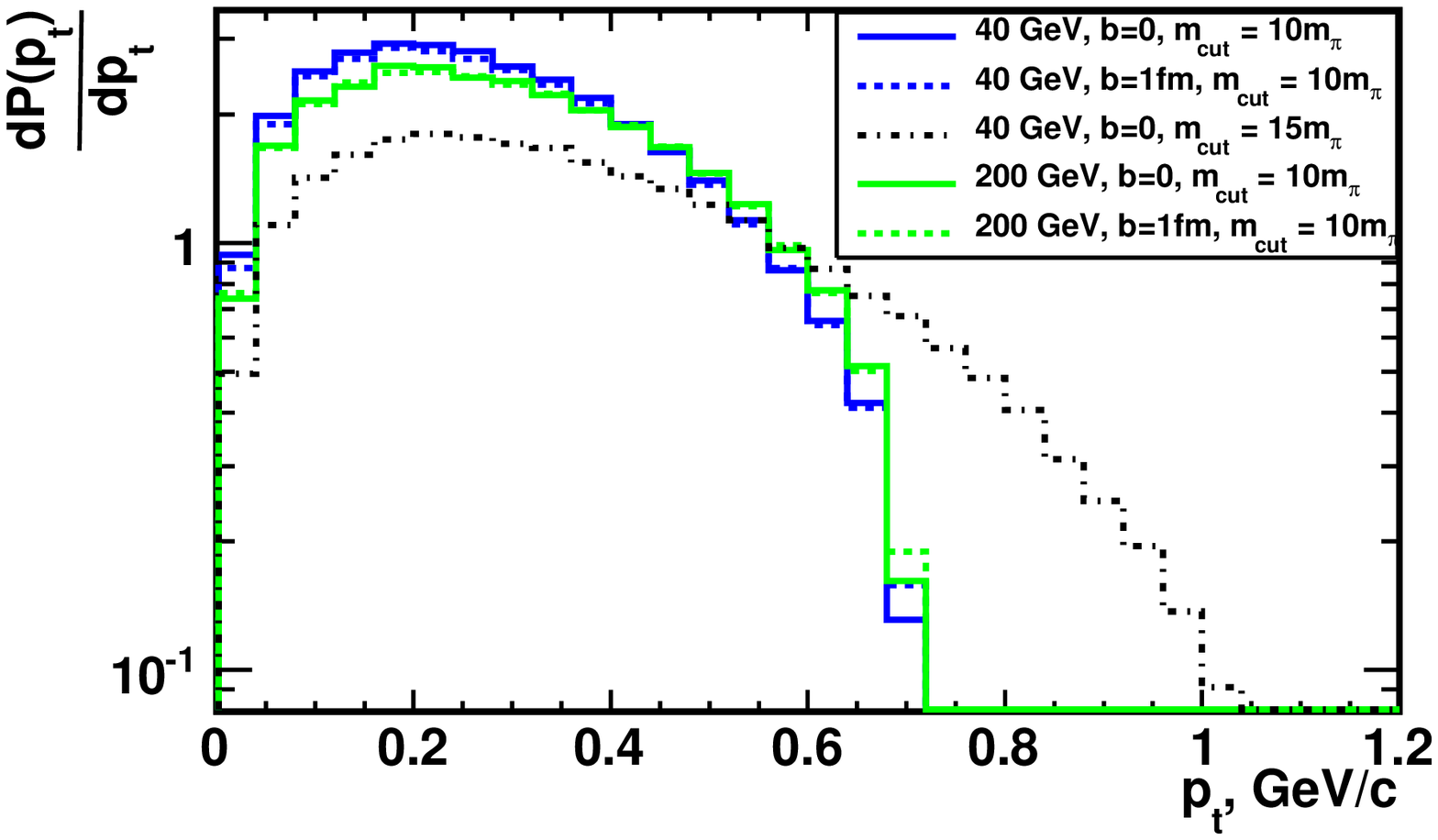}\hfill
\caption{Rapidity densities (left) and transverse momentum distributions(right) in our fragmentation scheme with 'invariant-type' initial conditions for different values of impact parameter, string c.m. energies and cutoff mass $m_{\rm cut}$.} \label{toyfrag:fig}
\end{figure}

In our numerical estimates we fix string tension and decay parameter at the level of $$
\varkappa = 1~\text{GeV/fm}, \quad k = 1~\text{GeV/fm}^2$$ and cutoff mass at $m_{\rm cut} = 10 m_\pi$
as exact fitting of experimental data on multiplicity and transverse momentum goes beyond the scope of this work. This set of parameters provides reasonable rapidity distribution of average multiplicity with plato at zero rapidity and $\left. dN/dy\right|_{y=0} \approx 2$ (which corresponds to $\left.dN_{\rm charged}/dy\right|_{y=0} \approx 1.25$) and transverse momentum  distribution with  $\langle p_t \rangle = 0.28$~GeV almost independent on the total energy of the initial string, see fig.~\ref{toyfrag:fig}. The transverse momentum distribution drops to zero at approximately $0.5 m_{\rm cut}$~GeV as seen from the right panel of fig.~\ref{toyfrag:fig}, which is of course a consequence of the cutoff mass which we have in our toy fragmentation model. As one can see, impact parameter $b$ has a minor influence on these distributions.

Asimuthal anisotropy of the transverse momentum distribution is usually described by the Fourier coefficients in the decomposition of asimuthal spectrum \cite{Voloshin:1994mz,Poskanzer:1998yz}:
\begin{equation}
\frac{d\sigma}{d^2b dp_t^2 d\phi} \sim 1 + \sum_{n=1}^\infty 2v_n(b, p_t , y) \cos(n \phi) ,
\end{equation}
where angle $\phi$ is between the direction of transverse momentum and reaction plane which contains beam axis and impact parameter.
So the flow coefficients $v_n$,  provided the direction of impact parameter is known, can be evaluated as
\begin{equation}
v_n = \langle \cos\big[n(\widehat{{\bf p}_t {\bf b}})\big] \rangle
\end{equation}
with average taken over all particles in the sample.
The first $v_1$ and the second $v_2$ coefficients are known as directed and elliptic flow respectively.

\begin{figure}

\includegraphics[width=\hsize]{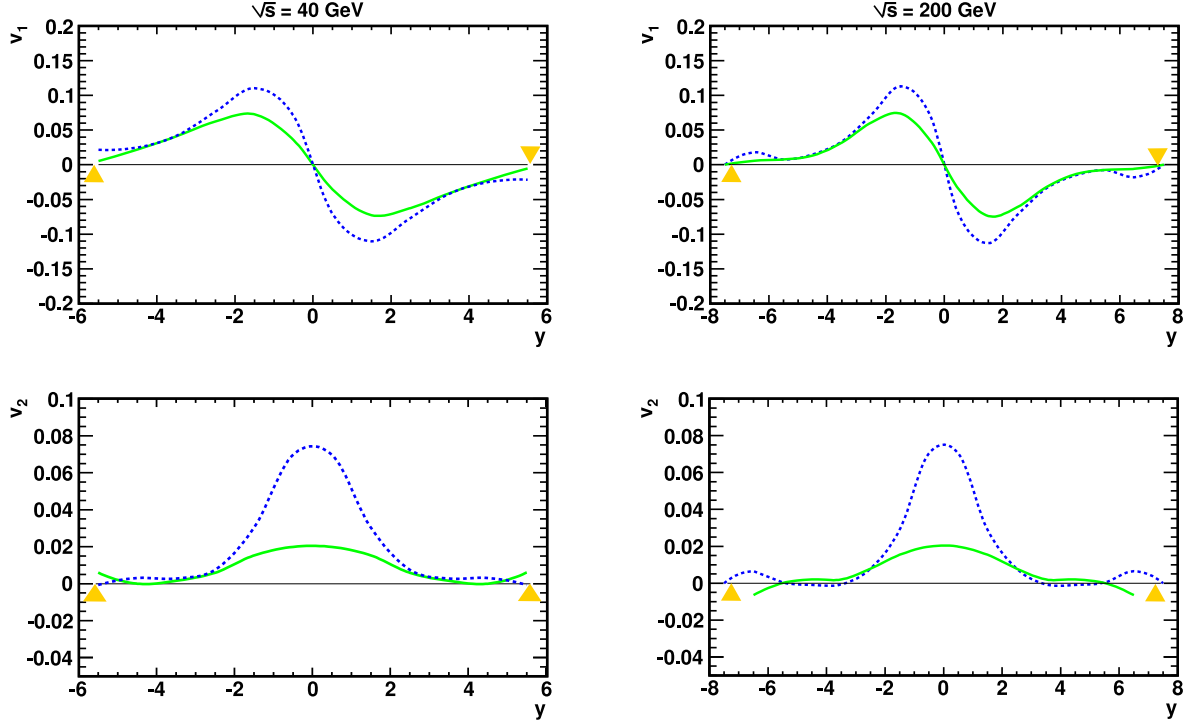}
\caption{Flow coefficients $v_1$ (upper plots) and $v_2$ (lower plots) for a decay of a single string with 'kink-type' initial conditions.  Left: c.m. string energy $\sqrt{s} = 40$~GeV; right: c.m. string energy $\sqrt{s} = 200$~GeV. Solid line~-- impact parameter $b = 0.5$~fm; dashed line~-- $ b=1 $~fm. Triangles denote edges of the rapidity distribution defined by $y_{\max}=\ln \sqrt{s}/m_{\pi}$}
 \label{flow-stick:fig} 
\end{figure}
\begin{figure}
 \includegraphics[width=\hsize]{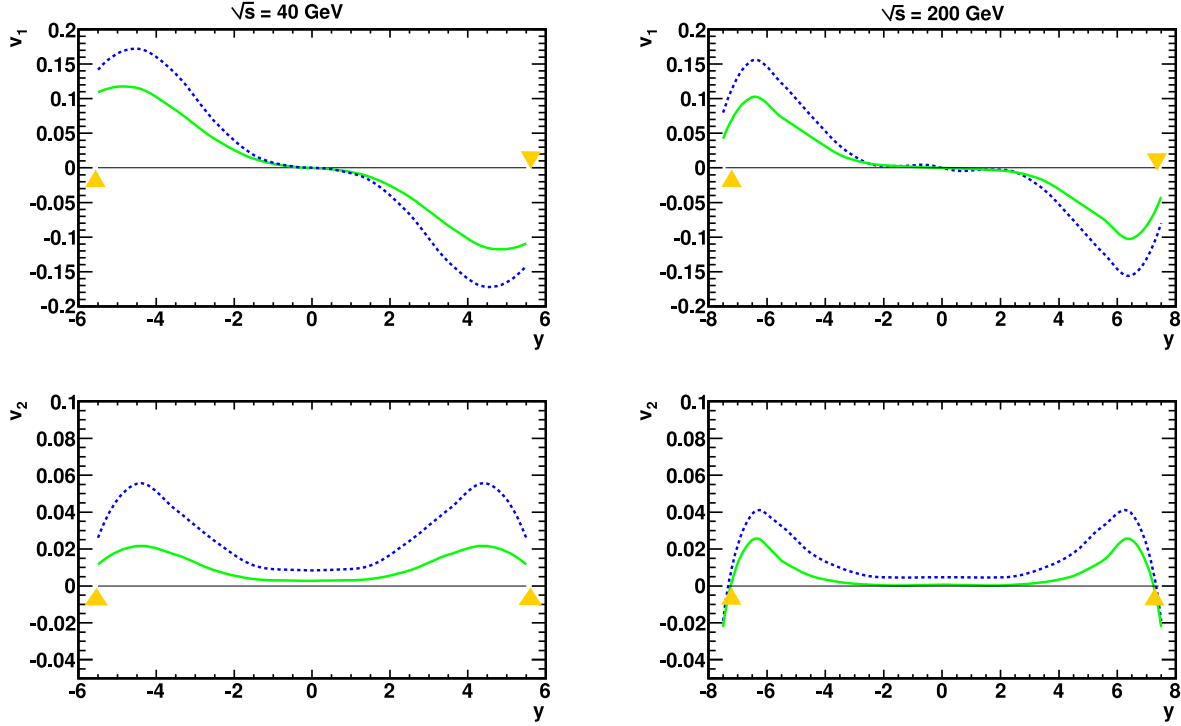}
\caption{The same as previous figure for 'invariant-type' conditions.}
 \label{flow-invar:fig}
\end{figure}

To explore parametric dependence of the flow induced by the two-dimensional string motion we perform computations of flow 
coefficients $v_1$ and $v_2$ for two types of initial conditions described in previous sections for different transverse separation between strings ends to which we attribute a role of impact parameter ${\bf b}$ and different string invariant masses (total energy in its c.m. frame). We dispose the impact parameter along the $x_2$ axis with initial momenta along $x_3$ and view string fragmentation from its center of mass frame. Hence, to evaluate flow coefficients integrated over $p_t$, which we do as a function of rapidity, we compute $\langle \cos \phi \rangle$ (for $v_1$) and $\langle \cos 2\phi $ (for $v_2$), where $\phi$ is the angle between transverse momentum of a particle and $x_2$ axis which direction coinsides with impact parameter $\bf b$ in our case.

\begin{figure}[!htb]
 \centering
\includegraphics[width=0.49\hsize]{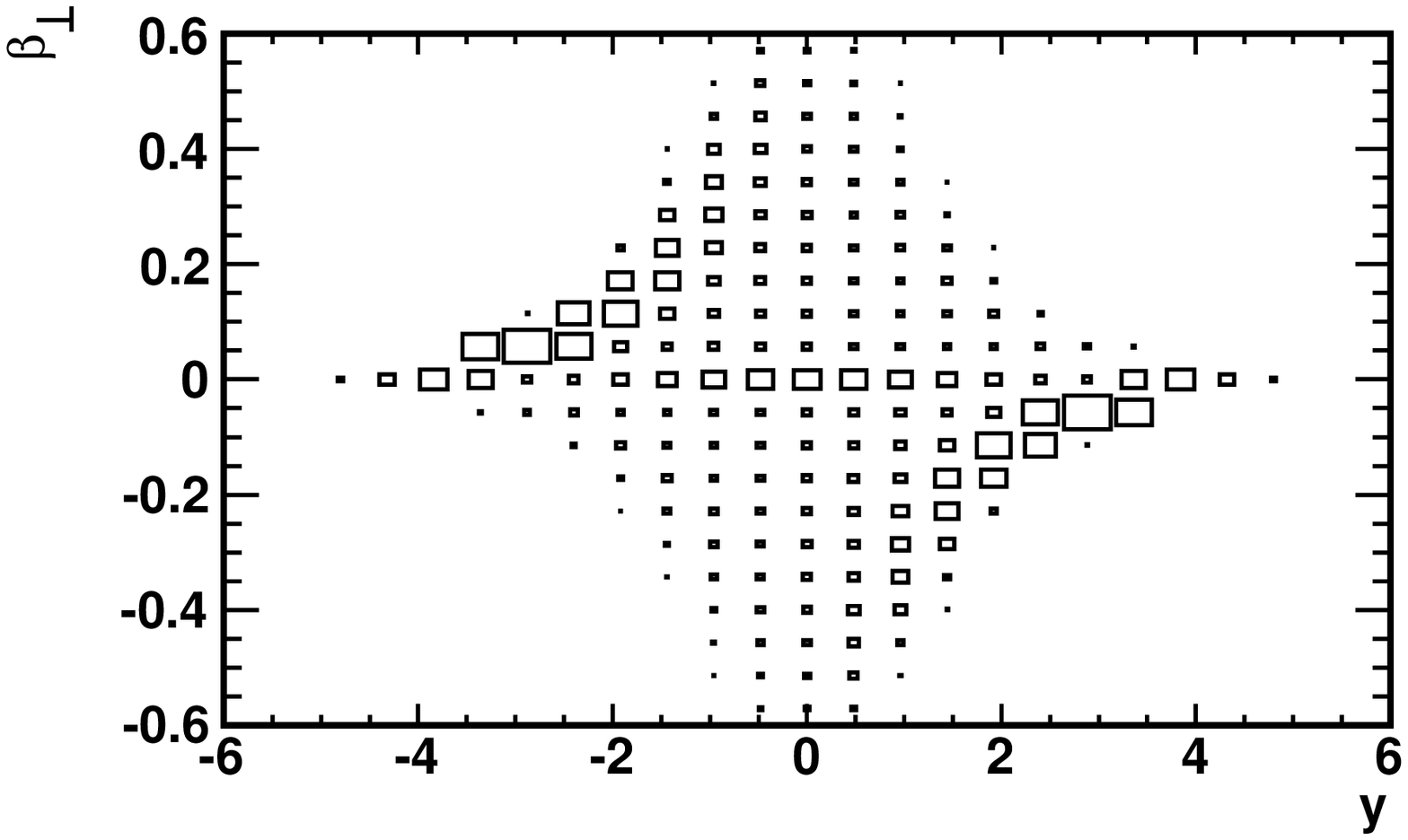}\includegraphics[width=0.49\hsize]{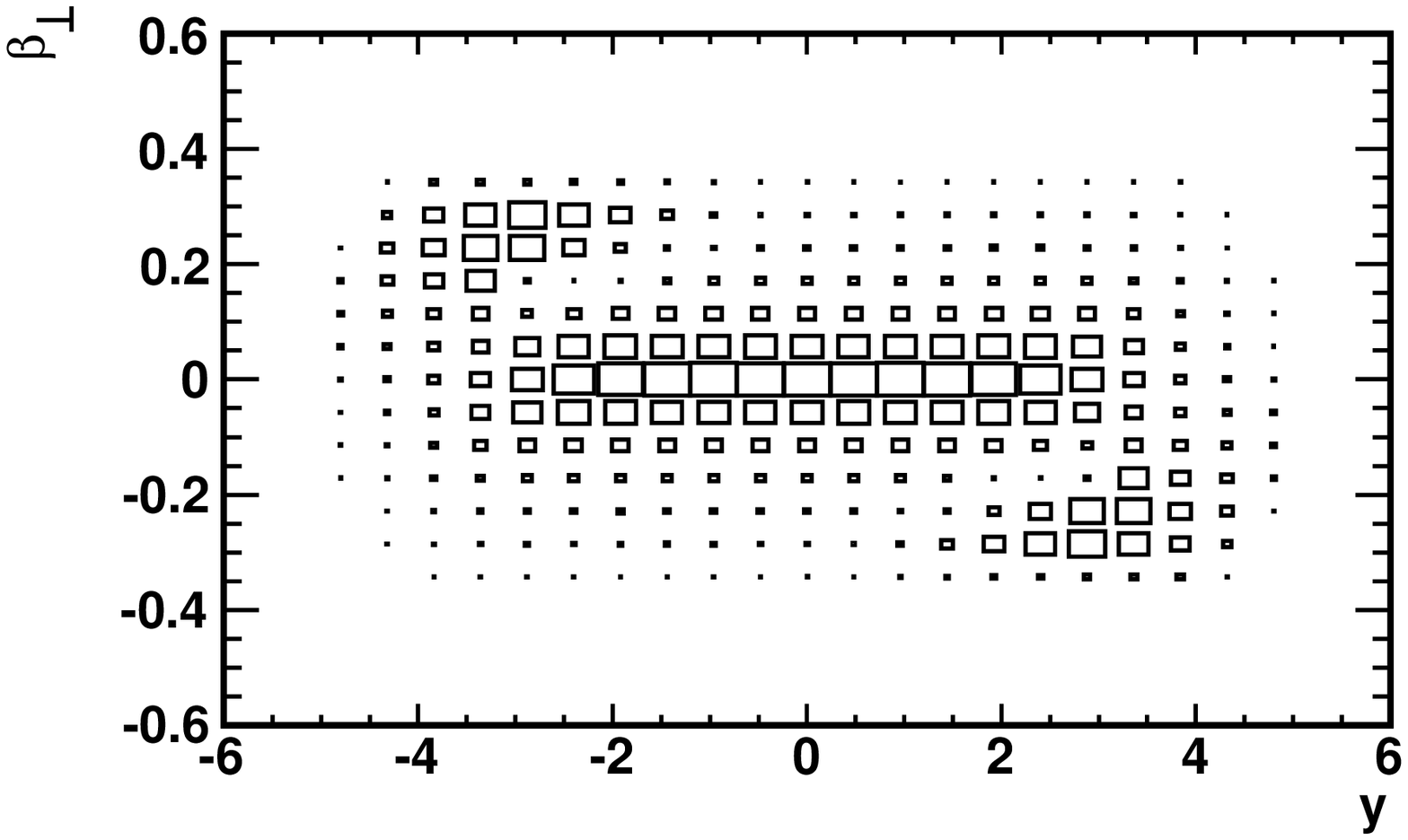} \label{betat:fig}
\caption{Rapidity and in-plane transverse velocity $\beta_\perp = v^{\text{fragment}}_\perp/c$ distribution of primary fragments: 2D plot, showing number of counts as a function of transverse velocity and rapidity $N(\beta_t,y)$. Surface of the box on the plot is proportionnal to bin content. Left: 'kink-type' initial conditions; right: 'invariant-type' initial conditions, string c.m. energy $\sqrt{s} = 40$ GeV.  Both cases are for impact parameter $b = 1$~fm. }
\end{figure}

Results of the computation are presented in two sets of figures. As observed, kink-type and invariant-type initial conditions produce different flow patterns. Some qualitative comments can be given on the basis of string motion depicted in fig.~\ref{string:figure:invar}. As suggested in  \cite{Sjostrand}, the breakup points are causally disconnected, which means that for (not too large) primary fragment their transverse momentum closely follows velocity of the adjacent breakpoints which cut this fragment away. In the 'kink-type' case, fragmentation starts in the region of a string which is stretched at zero time between two kinks and which has zero rapidity. The first break produces two pieces which have one of their ends slow. Subsequent breakpoints appear on the parts of the string which make 45$^\circ$ with the $x_3$ axis cutting off fragments from the slow ends of the two secondary strings. This populates rapidity distribution starting from the central region. In the beginning fragments with larger lengths are cut off, which hence carry larger momentum along $x_2$. So anisotropy in this case is stronger at central rapidity. 
In the case of 'invariant' type initial conditions in-plane transverse velocities of the inner parts of the string are much smaller than for the 'kink' case. However ends of the string acquire transverse velocities which leads to much more pronounced flow at forward and backward rapidities.
This is illustrated in figure~\ref{betat:fig} showing the in-plane transverse velocity distribution of primary fragments (produced in the model via classical string fragmentation as described in previous section) prior to final decay into two particles.

It is remarkable that the sign of the directed flow coefficient $v_1$ in both cases is opposite to the 'spectator flow', that is, sign of the directed flow is the same as in the experimentally observed directed flow for pi-mesons in Au-Au collisions~\cite{RHICv1}.

\section{Discussion and concluding remarks.}

String motion and fragmentation started from two initial conditions depicted here represent two in some sence extreme cases. For the 'kink-type' initial conditions almost all of the string's energy is initially carried by its ends while for the 'invariant-type' the energy is distributed in the interior of the string.
Hence the value for the elliptic flow at central rapidity obtained for the 'kink' case (fig.~\ref{flow-stick:fig}) should be rather viewed upon as the upper estimate for the part of the flow originating from the anisotropy of the quark-gluon string decay itself. In general it is remarkable that in both scenarios we obtain positive elliptic flow and the sign of the directed flow coincides with that experimentally observed.

What is also remarkable, it is the sensitivity of the rapidity dependence of the flow in the model to the initial momentum distribution carried by the interior of the string, as illustrated by different rapidity dependencies of the flow depicted in figs. \ref{flow-stick:fig},\,\ref{flow-invar:fig}. We  note, that this statement should be valid for any fragmentation scheme, e.g. Field-Feynman \cite{Field:1977fa}, since, as mentioned, break points on a string are causally disconnected.

Application of the developed technique to the interaction of hadrons or nuclei is still not straightforward as requires careful treatment of several important features, such as, e.g., fluctuations in number and invariant masses of strings, produced at the first stage of interaction. A careful treatment of initial conditions with paying special attention to longitudinal momentum distribution in the inner part of a string and its correlation with the transverse position is probably the most important part. The fragmentation model itselt may also require future refinements, such as implementation of transverse momentum of $q$-$\bar q$ pair produced at breakup of a string, which is for simplicity set zero in this work but is necessary for more realistic $p_t$ distributions of final particles and accurate study of the $p_t$ dependence of the flow. 

These options however lie beyond the scope of the article which has a qualitative character. Its aim is to suggest a new mechanism  which can give contribution to the flow in the models based on quark-gluon string picture of particle production. It also could be important for setting up initial conditions for hydrodynamic evolution. So we point out that transverse separation of string ends and correlation between rapidity and transverse position of string parts can play an important role.

\medskip

Author is deeply thankful to O.V.\,Kancheli, A.B.\,Kaidalov and K.G.\,Boreskov for permanent interest and to T.Sj{\o}strand for stimulating discussions. He is also grateful to M.A.\,Braun and V.V.\,Vechernin for critical remarks which helped to improve the presentation. Discussions and careful reading by L.V.\,Bravina and E.E.\,Zabrodin are also acknowledged.
The work was supported by NFR Project 185664/V30 and RFBR grants 09-02-01327-a  and
08-02-91004-CERN\_a.

\end{document}